\documentclass[preprint,amsmath,amssymb,aps]{revtex4}
\usepackage{graphicx}
\usepackage{graphics}
\usepackage{amsmath}
\usepackage{dcolumn}
\usepackage{amssymb}
\usepackage{bm}
\usepackage[latin1]{inputenc}

\begin{document}
\title{Landau quantization for a neutral particle in presence of  topological defects}
\author{K. Bakke, L. R. Ribeiro, C. Furtado and J. R. Nascimento}
\email{kbakke,lrr,jroberto,furtado@fisica.ufpb.br} 
\affiliation{Departamento de F\'{\i}sica, Universidade Federal da Para\'{\i}ba, Caixa Postal 5008, 58051-970, Jo\~{a}o Pessoa, PB, Brazil}

\begin{abstract}
In this paper we study the Landau levels in the non-relativistic dynamics of a neutral particle which possesses a permanent magnetic dipole moment interacting with an external electric field in the curved spacetime background with the presence or absence of a torsion field. The eigenfunction and eigenvalues of Hamiltonian are obtained.  We  show that the presence of the topological defect  breaks the infinite degeneracy of the Landau levels arising in this system.  We also apply a duality transformation to discuss this same quantization for a dynamics of a neutral particle with a permanent electric dipole moment.

\end{abstract}
\maketitle

\section{Introduction}

One of the simplest possible physical problems is the quantum description of the motion of a charged particle in a two-dimensional space under an influence of a homogeneous magnetic field perpendicular to the velocity \cite{landau}. In this situation, Landau levels arise, that is, the system displays quantized energy levels in the plane perpendicular to the magnetic field. The Landau levels have a important role in the study of several physical problems, such as, \emph{e.g.}, the quantum Hall effect \cite{prange}, different two-dimensional surfaces \cite{comtet,grosche,dunne}, anyons' excitations in a rotating Bose--Einstein condensate \cite{paredes1,paredes2}, and other ones.  The investigation of appearance of a quantum phases in the quantum dynamics of the electromagnetic dipole is based on the Aharonov-Casher effect \cite{ac} for the dynamics of magnetic dipole submitted an external  electric field, that is reciprocal effect of the Aharonov-Bohm \cite{pr:ab} effect. The dual of Aharonov-Casher effect was studied independently by He and McKellar \cite{hm} and Wilkens \cite{w}, who demonstrated  that  the quantum dynamics of the electric dipole in the presence of a magnetic field exhibits a geometrical quantum phase. In fact, the dynamics of dipoles can give rise to a variety of interesting physical effects \cite{spa,pac,prafurt}. Ericsson and Sj\"oqvist developed an analog of Landau quantization for neutral particles in the presence of an external electric field \cite{er}. The idea is based on the Aharonov--Casher effect within which neutral particles can interact with an electric field via a non-zero magnetic dipole moment. In the same way, the analog of Landau quantization for neutral particles possessing a non-zero electric dipole moment can be obtained by use of the He-McKellar-Wilkens effect \cite{2}. To solve the problem of magnetic monopoles in this system, we proposed the study of an analog of Landau quantization in a quantum dynamics of an induced electric dipole in the presence of crossed electric and magnetic fields \cite{3}. 

In last few decades, the subject of topological defects has drawn special attention in several areas of physics \cite{vil,kibble,kleinert,anp:dia,kat,furt}. Recently, the quantum dynamics of relativistic \cite{jackiw,vgeuhidro,euge,furtassi,josevi} and non-relativistic \cite{placlau,baus,epl0,azev} particles in the presence of a topological defects has been studied.
The  influence of a topological defect to the Landau levels in the presence of a topological defect has been investigated in recent years \cite{furtpla1,furtepl1}. In \cite{1}, the Landau levels have been investigated in the continuum elastic medium with topological defect in  the presence of an external magnetic field. It was shown that the presence of topological defects breaks the infinite degeneracy of the Landau levels. In \cite{4}, the Landau levels were investigated in the presence of a density of screw dislocations. 

The relativistic and non-relativistic quantum dynamics of a neutral particle with permanent magnetic and electric dipole moments which interacts with externals fields was studied in flat spacetime in \cite{anan}, in the curved spacetime in \cite{bf1,euge} and in the presence of a torsion field in \cite{bf2}.  In this paper, we construct an analog of Landau quantization for a neutral particle with permanent magnetic dipole moment which interacts with an external electric field in a cosmic string and cosmic dislocation spacetime.  

This paper is organized as follows. In section II we study the analog of Landau levels obtained in the cosmic string background. In section III, we analyze the Landau Levels in flat spacetime, but in the presence of a torsion. In section IV, we discuss the Landau levels in curved spacetime background with the presence of the torsion. In section V, we present our conclusions.

\section{Landau Level for a neutral particle in the cosmic string spacetime}\label{sectionII}

In this section, we  study the Landau levels which arise within the dynamics of a neutral particle with nonzero magnetic dipole in the presence of a topological defect, that is, a cosmic string. The line element  in the curved spacetime with such a topological defect is given by the following expression \cite{vil}
\begin{eqnarray}
ds^{2}=-dt^{2}+d\rho^{2}+\eta^{2}\rho^{2}d\varphi^{2}+dz^{2}.
\label{2.1}
\end{eqnarray}
where  $\eta$ is called deficit angle and is defined as $\eta=1-4\varpi G/c^{2}$ where  $\varpi$ is the linear mass density of the cosmic string. The azimuthal angle varies in the interval: $0\leq\varphi<2\pi$. The deficit angle can assume only values for which $\eta<1$ (unlike of this, in ~\cite{kat,furt}, it can assume values greater than 1, which correspond to an anti-conical space-time with negative curvature). This geometry possesses a conical singularity represented by the following curvature tensor
\begin{eqnarray}
\label{curv}\label{curva}
R_{\rho,\varphi}^{\rho,\varphi}=\frac{1-\eta}{4\eta}\delta_{2}(\vec{r}),
\end{eqnarray}
where $\delta_{2}(\vec{r})$ is the two-dimensional delta function. This behavior of the curvature tensor is denominated as conical singularity~\cite{staro}. The conical singularity gives rise to the curvature concentrated on the cosmic string axis, in all  other places the curvature is zero.

Through the study of the relativistic dynamics of this neutral particle, it is convenient to construct a  frame which allows us to define the spinors in the curved spacetime. We can introduce the frame using a non-coordinate basis $\hat{\theta}^{a}=e^{a}_{\,\,\,\mu}\,dx^{\mu}$, with its components $e^{a}_{\,\,\,\mu}\left(x\right)$ satisfy the following relation \cite{bd,naka}
\begin{eqnarray}
g_{\mu\nu}\left(x\right)=e^{a}_{\,\,\,\mu}\left(x\right)\,e^{b}_{\,\,\,\nu}\left(x\right)\,\eta_{ab}.
\label{2.2}
\end{eqnarray}
The components of the non-coordinate basis $e^{a}_{\,\,\,\mu}\left(x\right)$ form \textit{tetrad or Vierbein}. The tetrad has an inverse one defined as $dx^{\mu}=e^{\mu}_{\,\,\,a}\,\hat{\theta}^{a}$, where 
\begin{eqnarray}
e^{a}_{\,\,\,\mu}\,e^{\mu}_{\,\,\,b}=\delta^{a}_{\,\,\,b},\,\,\,\,\,\,\,e^{\mu}_{\,\,\,a}\,e^{a}_{\,\,\,\nu}=\delta^{\mu}_{\,\,\,\nu}.
\label{2.3a}
\end{eqnarray}
For the metric corresponding to a cosmic string we choose the tetrad and its inverse to be
\begin{eqnarray}
e^{a}_{\,\,\,\mu}=\left(
\begin{array}{cccc}
1 & 0 & 0 & 0 \\
0 & \cos\varphi & -\eta\rho\sin\varphi & 0 \\
0 & \sin\varphi & \eta\rho\cos\varphi & 0 \\
0 & 0 & 0 & 1 \\
\end{array}\right),\,\,\,\,\,\,\,\,\,\,\,\,\,e^{\mu}_{\,\,\,a}=\left(
\begin{array}{cccc}
1 & 0 & 0 & 0 \\
0 & \cos\varphi & \sin\varphi & 0 \\
0 & -\frac{\sin\varphi}{\eta\rho} & \frac{\cos\varphi}{\eta\rho} & 0 \\
0 & 0 & 0 & 1 \\
\end{array}\right),
\label{2.3}
\end{eqnarray}
which yields the correct flat spacetime limit for $\eta=1$. Taking the Cartan's structure equations
\begin{eqnarray}
T^{a}=d\hat{\theta}^{a}+\omega^{a}_{\,\,\,b}\,\hat{\theta}^{b},
\label{2.3a}
\end{eqnarray}
with $T^{a}=T^{a}_{\,\,\,\mu\nu}\,dx^{\mu}\,dx^{\nu}$ and $\omega^{a}_{\,\,\,b}=\omega_{\mu\,\,\,b}^{\,\,\,a}$. Solving for the tetrads given in (\ref{2.3}), we obtain
\begin{eqnarray}
\omega_{\varphi\,\,\,2}^{\,\,\,1}=-\omega_{\varphi\,\,\,1}^{\,\,\,2}=1-\eta.
\label{2.3b}
\end{eqnarray}

The relativistic dynamics of the neutral particle in this curved spacetime was studied in \cite{bf1}. In the same paper, the non-relativistic behavior of the neutral particle in curved spacetime was obtained through the application of the Foldy-Wouthuysen approximation \cite{fw} to the Dirac equation. We suggest that the dipole magnetic moments are parallel to the $z$-axis of the spacetime. The non-relativistic equation is
\begin{eqnarray}
i\frac{\partial\Psi}{\partial t}&=&m\Psi+\left[\frac{1}{2m}\left(\vec{p}+\vec{\Xi}\right)^{2}-\frac{\mu^{2}\,E^{2}}{2m}+\frac{\mu}{2m}\vec{\nabla}\cdot\vec{E}+\mu\hat{n}\cdot\vec{B}\right]\Psi,
\label{2.4}
\end{eqnarray}
where the first term above represents the rest energy of the neutral particle and the last four terms represents the Schr\"odinger-Pauli equation in curved spacetime. The unit vector $\hat{n}$ indicates the direction of the magnetic dipole moment and the vector $\vec{\Xi}$ was defined in such a way that its components are given in the local reference frame by
\begin{eqnarray}
\Xi_{j}=\mu\,(\hat{n}\times\vec{E})_{j}+\frac{1}{2}\left(1-\eta\right)\,e^{\varphi}_{\,\,\,j}.
\label{2.5}
\end{eqnarray}
Notice that the first term in (\ref{2.5}) is the Aharonov-Casher coupling and the another term arises due to the presence of a topological defect. Here we employ the same procedure which was adopted by Ericsson and Sj\"oqvist \cite{er} in the flat case. We need a field configuration in which that quantum dynamics of  magnetic dipole would exhibit bound states.

At this moment we must configure the electric field in a way providing the conditions pointed out in the reference \cite{er} to be satisfied. Thus, we choose that the electric field is determined by the only non-zero component, being of the form
\begin{eqnarray}
\vec{E}=\frac{\lambda\,\rho}{2}\,\hat{\rho},
\label{2.6}
\end{eqnarray}
with $\lambda$ is a linear density charge. For this field configuration is easy to verify that the conditions \cite{er}
\begin{eqnarray}
\frac{\partial\vec{E}}{\partial t}=0;\,\,\,\,\,\,\,\,\,\,\vec{\nabla}\times\vec{E}=0,
\label{2.7}
\end{eqnarray}
are satisfied. In this case, the Aharonov-Casher uniform magnetic field in the curved spacetime is given by 
\begin{eqnarray}
\vec{B}_{AC}=\vec{\nabla}\times\vec{\Xi}=\mu\,\lambda\,\hat{z}.
\label{2.8}
\end{eqnarray}
Using the field configuration given in the expression (\ref{2.6}), we  study the Landau-Aharonov-Casher quantization which occurs within the non-relativistic behavior of the neutral particle which has a permanent dipole moment. We apply the Schr\"odinger-Pauli equation (\ref{2.4}), and using the ansatz $\Psi=e^{-i\mathcal{E}t}\psi$ we have
\begin{eqnarray}
\mathcal{E}\psi&=&-\frac{1}{2m}\left[\frac{\partial^{2}\psi}{\partial\rho^{2}}+\frac{1}{\rho}\frac{\partial\psi}{\partial\rho}+\frac{1}{\eta^{2}\rho^{2}}\frac{\partial^{2}\psi}{\partial\varphi^{2}}+\frac{\partial^{2}\psi}{\partial z^{2}}\right]-\frac{i}{2m}\left(\frac{\mu\lambda}{\eta}+\frac{\left(1-\eta\right)}{\eta^{2}\rho^{2}}\right)\,\frac{\partial\psi}{\partial\varphi}\nonumber\\
&+&\frac{\mu^{2}\lambda^{2}}{8m}\,\rho^{2}\,\psi+\frac{1}{8m}\,\frac{\left(1-\eta\right)^{2}}{\eta^{2}\rho^{2}}\,\psi+\frac{\mu\lambda}{2m}\psi+\frac{\mu\lambda}{4m}\frac{\left(1-\eta\right)}{\eta}\psi.
\label{2.9}
\end{eqnarray}
The solution of the Schr\"odinger-Pauli equation  can be written in the form
\begin{eqnarray}
\psi\left(\rho,\varphi,z\right)=e^{il\varphi}\,e^{ikz}\,R\left(\rho\right).
\label{2.10}
\end{eqnarray}
Thus, the equation (\ref{2.9}) becomes 
\begin{eqnarray}
\mathcal{E}\,R=-\frac{1}{2m}\left(R''+\frac{1}{\rho}R'\right)+\frac{k^{2}}{2m}\,R+\frac{\gamma^{2}}{2m}\,\frac{R}{\eta^{2}\rho^{2}}+\frac{\gamma\omega}{2\eta}\,R+\frac{m\omega^{2}}{8}\,\rho^{2}\,R+\frac{\omega}{2}\,R,
\label{2.11}
\end{eqnarray}
where we defined $\gamma=l+\frac{\left(1-\eta\right)}{2}$ and $\omega=\mu\lambda/m$. Now, we make a convenient change of variables:
\begin{eqnarray}
\xi=\frac{m\omega}{2}\,\rho^{2},
\label{2.12}
\end{eqnarray} 
which gives us the expression
\begin{eqnarray}
\xi\,R''+R'+\left(\beta-\frac{\gamma^{2}}{4\eta^{2}\xi}-\frac{\xi}{4}\right)\,R=0,
\label{2.13}
\end{eqnarray}
where we defined  $\beta$ as
\begin{eqnarray}
\beta=\frac{\mathcal{E}}{\omega}-\frac{k^{2}}{2m\omega}-\frac{\gamma}{2\eta}-\frac{1}{2}.
\label{2.14}
\end{eqnarray}
We can write the solution of the Eq. (\ref{2.13}) as
\begin{eqnarray}
R\left(\xi\right)=e^{-\frac{\xi}{2}}\,\xi^{\frac{\left|\gamma\right|}{2\eta}}\,F\left(-\nu,\frac{\left|\gamma\right|}{\eta}+1;\xi\right).
\label{2.15}
\end{eqnarray}
This solution satisfies the usual asymptotic requirements and the finiteness at the origin for the bound state. We have the following equation for $F\left(-\nu,\frac{\left|\gamma\right|}{\eta}+1;\xi\right)$:
\begin{equation}\label{2.15a}
\xi F'' +\left(\frac{\left|\gamma\right|}{\eta} +1 -\xi\right)F'+ \left(\beta -\frac{1}{2} -\frac{\left|\gamma\right|}{2\eta}\right)F=0
\end{equation}
We find that the solution of equation (\ref{2.15a}) is the degenerated hypergeometric function
 \begin{eqnarray}
\label{hyper}
F=F\left[-\nu,\frac{\left|\gamma\right|}{\eta}+1;\xi\right]\;.
\end{eqnarray} 
The wave function is normalized if and only if the series in (\ref{hyper}) is a polynomial of degree $\nu$, therefore,
\begin{equation}
	\beta -\frac{1}{2} -\frac{\left|\gamma\right|}{2\eta}=\nu,
\end{equation}
where $\nu$ is an integer number. With this condition, we obtain discrete values for the energy, given by
\begin{eqnarray}
\mathcal{E}_{\nu,l}=\omega\left(\nu+\frac{\left|l+\frac{\left(1-\eta\right)}{2}\right|}{2\eta}+\frac{l+\frac{\left(1-\eta\right)}{2}}{2\eta}+1\right)+\frac{k^{2}}{2m},
\label{2.16}
\end{eqnarray}
with $\nu=0,1,2,...$ and $l=0,\pm1,\pm2,...$. These energy levels are the  Landau-Aharonov-Casher levels for a neutral particle with a permanent magnetic dipole moment which interacts with an external electric field in the presence of a topological defect. We can see clearly that, as in the reference \cite{1}, the topology of the defect breaks the infinite degeneracy of the Landau levels obtained in \cite{2} since the term $1/\eta$ does not be an integer. One should notice that these results are similar to results found for Landau levels of a charged particle in the presence of disclination \cite{furtpla1}. These levels are independent of the orbit center of a classical cyclotron motion and agree with the results of \cite{er,1,2,3}. However, they do not depend of the direction of the rotation as in \cite{2} due the definition of the angular frequency $\omega$ done for us above. In the limit $\eta\rightarrow 1$, we recuperate the energy levels obtained in \cite{2} plus the contribution of the free motion along the $z$-axis.

\section{Landau levels for a neutral particle in the presence of a cosmic dislocation}

In this section we investigate the Landau-Aharonov-Casher levels in the presence of a cosmic dislocation, with the line element is given by \cite{GL}
\begin{eqnarray}
ds^{2}=-dt^{2}+d\rho^{2}+\rho^{2}d\varphi^{2}+\left(dz-\chi\,d\varphi\right)^{2},
\label{3.1}
\end{eqnarray}
where the parameter $\chi$ is related with the torsion of the defect, or, within the crystallography language, with the Burgers vector. In the same way as in the previous section, it is convenient to construct a local reference frame with the spinor are defined in this background \cite{bd, naka}. We choose for our local vierbein \cite{bf2}
\begin{eqnarray}
e^{a}_{\,\,\,\mu}\left(x\right)=\left(
\begin{array}{cccc}
1 & 0 & 0 & 0 \\
0 & \cos\varphi & -\rho\sin\varphi & 0 \\
0 & \sin\varphi & \rho\cos\varphi & 0 \\
0 & 0 & \chi & 1 \\
\end{array}\right),\,\,\,
e^{\mu}_{\,\,\,a}\left(x\right)=\left(
\begin{array}{cccc}
1 & 0 & 0 & 0 \\
0 & \cos\varphi & \sin\varphi & 0 \\
0 & -\frac{\sin\varphi}{\rho} & \frac{\cos\varphi}{\rho} & 0 \\
0 & \frac{\chi}{\rho}\sin\varphi & -\frac{\chi}{\rho}\cos\varphi & 1 \\
\end{array}\right).
\label{3.2}
\end{eqnarray}

Solving the Cartan's structure equation (\ref{2.3a}), we have that the contribution given by the torsion field is
\begin{eqnarray}
T^{3}=\chi\,\delta\left(\rho\right)\,d\varphi\wedge d\rho.
\label{3.2b}
\end{eqnarray}

Again we suggest that the directions of the magnetic moments of dipoles are parallel to the $z$-axis of the spacetime. Following the reference \cite{bf2}, we conclude that the non-relativistic behavior of the neutral particle in the flat spacetime but in the presence of the torsion is given by the Schr\"odinger-Pauli equation
\begin{eqnarray}
\mathcal{E}\psi&=&\frac{1}{2m}\left(\vec{p}+\vec{\Xi}\right)^{2}-\frac{\mu^{2}\,E^{2}}{2m}+\frac{1}{8}\,\mu\,\hat{n}\cdot\vec{S}+\frac{\mu}{2m}\vec{\nabla}\cdot\vec{E}+\mu\,\hat{n}\cdot\vec{B},
\label{3.3}
\end{eqnarray}
with the vector $\vec{\Xi}$ has the following components
\begin{eqnarray}
\Xi_{i}=\mu\,\left(\hat{n}\times\vec{E}\right)_{i}-\frac{1}{8}\,S^{0}\,n_{i},
\label{3.4}
\end{eqnarray}
and we write them in terms of the irreducible component of the torsion tensor $S^{\nu}=\epsilon^{\mu\alpha\beta\gamma}\,T_{\alpha\beta\gamma}$, called axial $4$-vector.

Notice that the first term in Eq.(\ref{3.4}) is an Aharonov-Casher coupling and the second one is related with the torsion of topological defect.
To define the field configurations in this background, we suggest that the dipole moments are parallel to the $z$-axis again and that the electric field is outside of the defect. Thus, the electric field satisfying the conditions pointed out in the reference \cite{er} is given by the expression (\ref{2.6}), with the Aharonov-Casher uniform magnetic field is given by (\ref{2.8}). In this case, the Schr\"odinger-Pauli equation in the presence of the torsion field takes the form
\begin{eqnarray}
\mathcal{E}\psi&=&-\frac{1}{2m}\left[\frac{\partial^{2}\psi}{\partial\rho^{2}}+\frac{1}{\rho}\frac{\partial\psi}{\partial\rho}+\frac{1}{\rho^{2}}\left(\frac{\partial}{\partial\varphi}-\chi\frac{\partial}{\partial z}\right)^{2}\psi+\frac{\partial^{2}\psi}{\partial z^{2}}\right]-\frac{i}{2m}\,\mu\,\lambda\,\left(\frac{\partial}{\partial\varphi}-\chi\frac{\partial}{\partial z}\right)\psi\nonumber\\
&+&\frac{\mu^{2}\lambda^{2}}{8m}\,\rho^{2}\,\psi+\frac{\mu\lambda}{2m}\psi.
\label{3.5}
\end{eqnarray}
The solution of the above equation has the same general form (\ref{2.10}). Thus, we define $\omega=\mu\lambda/m$ and have
\begin{eqnarray}
\mathcal{E}\,R&=&-\frac{1}{2m}\left(R''+\frac{1}{\rho}R'\right)+\frac{k^{2}}{2m}\,R+\frac{\left(l-\chi\, k\right)^{2}}{2m}\,\frac{R}{\rho^{2}}\nonumber\\
&+&\frac{\omega}{2}\left(l-\chi\,k\right)\,R+\frac{m\omega^{2}}{8}\,\rho^{2}\,R+\frac{\omega}{2}\,R.
\label{3.6}
\end{eqnarray}
Again, we carry out a change of variables identical to (\ref{2.12}). Afterwards, we arrive at the equation
\begin{eqnarray}
\xi\,R''+R'+\left(\beta'-\frac{\left(l-\chi\, k\right)^{2}}{4\xi}-\frac{\xi}{4}\right)\,R=0,
\label{3.7}
\end{eqnarray}
where $\beta'$ is
\begin{eqnarray}
\beta'=\frac{\mathcal{E}}{\omega}-\frac{k^{2}}{2m\omega}-\frac{1}{2}\left(l-\chi\,k\right)-\frac{1}{2}.
\label{3.8}
\end{eqnarray}
To obtain the solution for the equation (\ref{3.7}) we use again the procedure developed in the section \ref{sectionII}, with the radial wavefunction is assumed to be of the following form
\begin{eqnarray}
R\left(\xi\right)=e^{-\frac{\xi}{2}}\,\xi^{\frac{\left|l-\chi\,k\right|}{2}}\,F\left(-\nu,\left|l-\chi\,k\right|+1;\xi\right).
\label{3.9}
\end{eqnarray}

The energy levels in this case look like
\begin{eqnarray}
\mathcal{E}_{\nu,l}=\omega\left(\nu+\frac{\left|l-\chi\,k\right|}{2}+\frac{\left(l-\chi\,k\right)}{2}+1\right)+\frac{k^{2}}{2m},
\label{3.10}
\end{eqnarray}
with $\nu=0,1,2,...$ and $l=0,\pm1,\pm2,...$. This expression is the analog of the Landau levels of a neutral particle with a permanent magnetic dipole moment interacting with an external electric field in the presence of defect, more precisely, in the presence of a torsion field plus a contribution of the kinetic energy of the free motion in the $z$-direction of the spacetime. Notice that this result is similar to the result obtained for Landau levels in the presence of a screw dislocation in \cite{furtepl1}. Again, the infinite degeneracy of the Landau levels obtained in \cite{2} is eliminated due the coupling of the torsion with the angular moment $l$, which agrees with the results given in \cite{1}. However, if we take $\chi=0$ we recuperate the infinite degeneracy pointed out in \cite{2}.

\section{Landau levels for the neutral particle in the presence of the massive cosmic dislocation}

In this section we study the Landau-Aharonov-Casher quantization in the quantum dynamics of a neutral particle with a permanent magnetic dipole moment in the presence of a massive cosmic string \cite{GL}, with the line element is given by
\begin{eqnarray}
ds^{2}=-dt^{2}+d\rho^{2}+\eta^{2}\rho^{2}d\varphi^{2}+\left(dz-\chi\,d\varphi\right)^{2},
\label{4.1}
\end{eqnarray}

The matrix form of the tetrad and its inverse is
\begin{eqnarray}
e^{a}_{\,\,\,\mu}\left(x\right)=\left(
\begin{array}{cccc}
1 & 0 & 0 & 0 \\
0 & \cos\varphi & -\eta\rho\sin\varphi & 0 \\
0 & \sin\varphi & \eta\rho\cos\varphi & 0 \\
0 & 0 & \chi & 1 \\
\end{array}\right),\,\,\,
e^{\mu}_{\,\,\,a}\left(x\right)=\left(
\begin{array}{cccc}
1 & 0 & 0 & 0 \\
0 & \cos\varphi & \sin\varphi & 0 \\
0 & -\frac{\sin\varphi}{\eta\rho} & \frac{\cos\varphi}{\eta\rho} & 0 \\
0 & \frac{\chi}{\eta\rho}\sin\varphi & -\frac{\chi}{\eta\rho}\cos\varphi & 1 \\
\end{array}\right).
\end{eqnarray}

Following the study carried out in \cite{bf2}, we find that the non-relativistic behavior of the neutral particle in the flat spacetime but in the presence of the torsion is described by the Schr\"odinger-Pauli equation
\begin{eqnarray}
\mathcal{E}\psi&=&\frac{1}{2m}\left(\vec{p}+\vec{\Xi}\right)^{2}-\frac{\mu^{2}\,E^{2}}{2m}+\frac{1}{8}\,\mu\,\hat{n}\cdot\vec{S}+\frac{\mu}{2m}\vec{\nabla}\cdot\vec{E}+\mu\,\hat{n}\cdot\vec{B},
\label{4.3}
\end{eqnarray}
with the vector $\vec{\Xi}$ in this case has the following components
\begin{eqnarray}
\Xi_{i}=\mu\,\left(\hat{n}\times\vec{E}\right)_{i}+\frac{1}{2}\left(1-\eta\right)-\frac{1}{8}\,S^{0}\,n_{i}.
\label{4.4}
\end{eqnarray}

Again, we suppose that the direction of the dipole moments is parallel to the $z$-axis of the topological defect. The field configurations is identical to the previous sections, thus the electric field is given by the expression (\ref{2.6}). Hence, the Schr\"odinger-Pauli equation is written as
\begin{eqnarray}
\mathcal{E}\psi&=&-\frac{1}{2m}\left[\frac{\partial^{2}\psi}{\partial\rho^{2}}+\frac{1}{\rho}\frac{\partial\psi}{\partial\rho}+\frac{1}{\eta^{2}\rho^{2}}\left(\frac{\partial}{\partial\varphi}-\chi\frac{\partial}{\partial z}\right)^{2}+\frac{\partial^{2}\psi}{\partial z^{2}}\right]-\frac{i}{2m}\frac{\mu\lambda}{\eta}\left(\frac{\partial}{\partial\varphi}-\chi\frac{\partial}{\partial z}\right)\psi\nonumber\\
&-&\frac{i}{2m}\frac{\left(1-\eta\right)}{\eta^{2}\rho^{2}}\left(\frac{\partial\psi}{\partial\varphi}-\chi\frac{\partial\psi}{\partial z}\right)+\frac{\mu^{2}\lambda^{2}}{8m}\,\rho^{2}\,\psi+\frac{1}{8m}\,\frac{\left(1-\eta\right)^{2}}{\eta^{2}\rho^{2}}\,\psi+\frac{\mu\lambda}{2m}\psi+\frac{\mu\lambda}{4m}\frac{\left(1-\eta\right)}{\eta}\psi.
\label{4.5}
\end{eqnarray}

The general solution is described by the ansatz (\ref{2.10}). Again, we define $\omega=\mu\lambda/m$ and obtain
\begin{eqnarray}
\mathcal{E}\,R=-\frac{1}{2m}\left(R''+\frac{1}{\rho}R'\right)+\frac{k^{2}}{2m}\,R+\frac{1}{2m}\frac{\zeta^{2}}{\eta^{2}\rho^{2}}\,R+\frac{\omega\gamma}{2\eta}\,R+\frac{m\omega^{2}}{8}\,\rho^{2}\,R+\frac{\omega}{2}\,R,
\label{4.6}
\end{eqnarray}
where $\zeta$ is defined as
\begin{eqnarray}
\zeta=\left(l-\chi\,k\right)+\frac{1}{2}\left(1-\eta\right).
\label{4.7}
\end{eqnarray}

Making the change of variables similar to (\ref{2.12}), we get
\begin{eqnarray}
\xi\,R''+R'+\left(\tilde{\beta}-\frac{\zeta^{2}}{4\eta^{2}\xi}-\frac{\xi}{4}\right)\,R=0,
\label{4.13}
\end{eqnarray}
where the parameter $\tilde{\beta}$ is defined as
\begin{eqnarray}
\tilde{\beta}=\frac{\mathcal{E}}{\omega}-\frac{k^{2}}{2m\omega}-\frac{\zeta}{2\eta}-\frac{1}{2}.
\label{4.12}
\end{eqnarray}

The solution for this equation can be written in the form
\begin{eqnarray}
R\left(\xi\right)=e^{-\frac{\xi}{2}}\,\xi^{\frac{\left|\zeta\right|}{2\eta}}\,F\left(-\nu,\frac{\left|\zeta\right|}{\eta}+1;\xi\right).
\label{4.9}
\end{eqnarray}

The energy levels associated with this system are
\begin{eqnarray}
\mathcal{E}_{\nu,l}=\omega\left(\nu+\frac{\left|\left(l-\chi\,k\right)+\frac{1}{2}\left(1-\eta\right)\right|}{2\eta}+\frac{\left(l-\chi\,k\right)+\frac{1}{2}\left(1-\eta\right)}{2\eta}+1\right)+\frac{k^{2}}{2m},
\label{4.10}
\end{eqnarray}
with $\nu=0,1,2,...$ and $l,k=0,\pm1,\pm2,...$. This expression for the energy levels is also the analog of the Landau levels in the presence of the deficit angle and the the torsion. Note that if we consider $\chi=0$ we recuperate the Landau levels obtained in the expression (\ref{2.16}). However, if we consider $\eta\rightarrow 1$ we recuperate the Landau levels given in the expression (\ref{3.10}). In the same way as in the previous section, we find that the infinite degeneracy of the Landau levels obtained in \cite{2} is broken since the parameters of the topology $\eta$ and of the torsion $\chi$ in the expression (\ref{3.10}) are not integer. 

\section{conclusions}

In this paper we studied the quantum dynamics of a neutral particle with a permanent magnetic dipole moment in the presence of a topological defect. We have used the Aharonov-Casher coupling to obtain the analog of Landau quantization in this dynamics. We solved the Schr\"odinger-Pauli equation in the space-time of a cosmic string, cosmic dislocation an massiv cosmic dislocation. We have obtained the eigenvalues and eigenfunction in three cases. We demonstrated that in the appropriate limits we obtain the Landau-Aharonov-Casher levels found by Ericsson and Sj\"oqvist \cite{er}. We shown that the presence of defects breaks the infinite degeneracy of the Landau-Aharonov-Casher levels and observe that the topology of the spacetime breaks the infinite degeneracy of the Landau levels obtained in flat spacetime \cite{2} since the term $1/\eta$ is not an integer. In the same way, the presence of the torsion, in the cosmic dislocation case, also breaks the infinite degeneracy of the Landau levels. Both results agree with the conclusions of \cite{1} where an external uniform magnetic field is applied in the space with topological defects.  At the end of the day, one can verify that in the limit $\eta\rightarrow 1$ and $\chi=0$, the spacetime becomes flat, and we recuperate the Landau levels obtained in \cite{2} and, consequently, the infinite degeneracy of the energy levels. 

We claim that we can obtain the results of quantum dynamics of neutral particle with a permanent electric dipole moment interacting with an external magnetic field via the He-Mckellar-Wilkens coupling \cite{hm,w}. We carried out this study employing the duality transformation in the equations of motions of the Landau-Aharonov-Casher problem in the presence of a topological defect and obtained the He-Mckellar-Wilkens quantization for a neutral particle in the presence of a topological defect. The equations of motion for LHMW case has the same form as the equations for LAC case. Indeed, changing $\mu$ by $-d$ and $\mathbf{E}$ by $\mathbf{B}$, we obtain the equation of motion for the latter. In this sense we observed in curved case the same duality transformations which take place in flat space-time.

\acknowledgments{Authors are grateful to A. Yu. Petrov for some criticism on the manuscript. This work was partially supported by PRONEX/FAPESQ-PB, FINEP, CNPq and CAPES/PROCAD)}


\end{document}